# Universality at the onset of turbulence in shear flows

Alberto de Lozar and Björn Hof

*Max Planck Institute for Dynamics and Self-Organisation, Bunsenstrasse 10, 37073 Göttingen, Germany*

**The volatile transition from quiescent laminar to strongly fluctuating turbulent dynamics in shear flows remains only poorly understood despite its practical importance and more than a century of intense research. The theoretical understanding of the transition process has been complicated by the lack of a linear instability mechanism and additionally by a catastrophic collapse of turbulence which can occur after extremely long lifetimes. Turbulence close to onset is investigated experimentally in three different geometries: pipe, duct and channel flow. The reverse transition from turbulent to laminar flow is observed to be a general feature of shear flows. A critical point is uncovered at slightly higher flow rates, where the nature of the flow changes abruptly to the generic case of expanding turbulence. The critical exponent associated with this phase transition is found to be universal. This confirms a conjecture made over 20 years ago based on an analogy between fluid turbulence and discrete models in statistical physics.**

Comprising flows through blood vessels, over airplane wings, in astrophysical accretion discs and the earth's atmosphere, shear flows are relevant to many different disciplines and occur over a large breadth of scales. A common feature of these flows is that the laminar and the turbulent state often co-exist over significant parameter ranges[1,2]. Even if a critical point exists at which the laminar state becomes unstable this transition point can be bypassed and turbulence typically sets in at much lower parameter values than expected from linear stability theory[3,4]. The onset of turbulence is accompanied by a large increase in drag, mixing rates and heat transfer and it is hence crucial for many processes to decide if flows are likely to be laminar or turbulent.

At low Reynolds numbers, Re, (e.g. Re=UD/ν <1500 in pipes, where U is the mean velocity of the flow, D the pipe diameter and ν the kinematic viscosity) flows are always observed to be laminar in practice. At larger Re and if disturbances to the flow are sufficiently strong[2,5] , transition to turbulent motion occurs. This transition is governed by the non-linear growth of finite perturbations and therefore the Reynolds number at which it occurs cannot be predicted by classical (linear) stability analysis. Due to the strong dependence on the magnitude of the finite perturbations, transition points vary greatly even under well controlled laboratory conditions. In ordinary pipes at Re~2000 flows are found to be strongly intermittent with sudden changes between laminar and turbulent segments[1,6]. Here the dynamics are dominated by the co-existence of laminar regions and turbulent 'spots' (referred to as 'puffs' in pipe flow) which are localized and have a finite size[7]. Turbulent spots have been observed to decay after long times[8,9] following a memoryless process. Recent studies in pipe flow have shown that surprisingly the lifetimes of these turbulent spots remain finite for all Reynolds numbers accessible to lifetime studies[10,11,12]. The super-exponential scaling with Re observed, only approaches an infinite lifetime asymptotically confirming turbulence to be transient in this regime. For sufficiently larger Reynolds numbers, beyond the range of lifetime studies, the turbulent spots grow and eventually invade the entire surrounding fluid, eliminating all laminar flow.

While the speed of laminar turbulent interfaces for pipes is well documented in the literature[7,13,14] the nature of the transition between localized and expanding turbulence could not be clarified nor is it clear at what exact parameter value it occurs. In the following we first characterize the lifetime behaviour of localized turbulence for three of the basic shear flows: pipe, duct and channel flow. Secondly investigations are carried out at higher Reynolds numbers, where in all three flows a phase transition to globally expanding turbulence is uncovered suggesting a universal transition process.

The three configurations selected, namely pipe, duct and channel flow (see figure 1 for details) differ in important aspects of their stability behaviour. Whereas in pipe and duct flow the laminar state is stable for all Reynolds numbers, laminar channel flow becomes linearly unstable[4] at a finite Reynolds number (Re=5772). Duct flow on the other hand has the particular feature that the averaged turbulent flow is not unidirectional but features a pair of eddies in each corner[15].

**Results and Discussion**

Lifetime distributions for localized turbulence have been determined as described in the experimental methods section. The survival rates of turbulent spots for channel and duct flow are shown in figure 2A and compared to the pipe data obtained in the earlier study[10]. The distinct S-shape of these curves[11] infers that the decay of turbulence is memoryless and can be described by an exponential ansatz: $P(t,Re)=\exp[-(t-t_0)/\tau(Re)]$ , where $\tau^{-1}$ is the decay rate and $t_0$ is related to the time for a perturbation to develop into a spot. The Reynolds number dependence of the decay rate $\tau^{-1}$ as function of Re is shown in figure 2B. The collapse of the data sets measured for different times confirms the memoryless nature of turbulence. The lifetimes, $\tau$, for all three flows scale super-exponentially with Re and only asymptotically approach infinite. The best fit was obtained by $\tau^{-1} = \exp(-\exp(c[Re-Re_0]))$, where c and $R_0$ are fitting parameters. Hence the description of turbulence as a super transient state, suggested for pipes[10,11], is a general feature of shear flows. The values of $Re_0$ (=1526 in pipes, 1085 in channels and 1221 in ducts) can be considered as a lower cut off in Re below which lifetimes of the spots are insignificant ($\tau<e$) and thus self-sustained turbulence becomes impracticable. At the high Reynolds number end the super-exponential increase makes it impossible to extend lifetime measurements (either experimentally or computationally) to sufficiently larger parameter values and consequently any change in the nature of turbulence cannot be inferred from lifetime studies.

It is well known that at higher Re turbulence in shear flows does not remain localized and spots invade into the adjacent laminar fluid. However the nature of the transition and the exact parameter value where it occurs are unknown. The main problem in characterizing the transition stems from the fact that the turbulent spots typically travel downstream at an advection speed which is not a priori known and which changes with Reynolds number. Once the Reynolds number is increased and turbulence begins to grow, the exact value of the laminar turbulent interface velocity is masked by the unknown change in the structures advection speed and vice versa. We therefore argue that measuring the speeds of the interfaces at the front and back end of the spots, as carried out in many earlier studies (e.g. 7 or 14), is not the appropriate measure to determine the nature of the transition. Instead looking at the difference between the velocities of the front and rear interface automatically selects the correct reference frame, i.e. co-moving with the centre of the turbulent spot and the so determined expansion rate is independent of the advection speed of the turbulent structure. A further process masking the true nature of this transition arises slightly below the critical point. Here spots begin to split (Re≈2400 in pipes), i.e. although any specific spot remains localized it will eventually contaminate the adjacent laminar flow and seed a second spot some distance downstream. To separate this process from the transition to expanding turbulence we only focus on the growth rate of continuous turbulent regions. Measurements of the expansion speeds as a function of Re were carried out in the three facilities and the square of the growth rate, G, as function of Re is shown in figure 3. $G^2$ is initially zero (localized spots) and at some critical parameter value starts to increase linearly with Re marking a phase transition from localized to expanding turbulence. From the linear fit the critical Reynolds number can be obtained and is found to be $Re_c$=2550 for pipes, 1480 for channels and 2250 for ducts. Based on discrete models in statistical physics it had been conjectured[16] that such a phase transition should occur in shear flows and that close to the critical point speeds of laminar turbulent interfaces scale with $\varepsilon^{\alpha}$ where $\varepsilon = (Re - Re_c) / Re_c$ (see also references 17 and 18) and that the exponent may be related to critical exponents in directed percolation. While the relation between turbulent fronts and discrete system in statistical physics is far from obvious[19] we indeed observe the same exponent ($\alpha$=0.5) for all three systems. Square root dependence of front velocities on a parameter at a critical point has also been observed in reaction diffusion problems[20] and models in statistical physics[21,22].

Any predictions concerning lifetimes of turbulence cannot be extrapolated beyond the phase transition point. In principle segments of an expanding turbulent spot could still decay before being invaded by the adjacent turbulent flow again. Such measurements will be hard to achieve in pipes and ducts where (according to the observed scaling laws) lifetimes on the order of $10^{150}$ and $10^{230}$ years are expected at the respective phase transition points. In channels remarkably the phase transition sets in much earlier and the required observation time here is only 2.5 hours which can be realized experimentally (though it is just out of reach of the present facility).

**Conclusions**

The nature of turbulence close to onset has been characterized for three different flows. Although the studied flows differ in many aspects the transition scenario is qualitatively identical. Universality close to a critical point has been observed in many complex systems, but prior to this study not for turbulent flows. The scaling behaviour uncovered is likely to inspire new theoretical approaches and to significantly improve our understanding of one of the most complex phenomena in nature.

**Methods**

For the lifetime studies the experimental procedure was identical to that described in references 10 and 11. As in the earlier study for pipe flow first a perturbation was applied at a fixed position upstream of the channel /duct outlet. The perturbation amplitude was chosen large enough to trigger transition to turbulence and the duration of the perturbations was set to between 10 and 20 D/U ensuring that only a single turbulent spot nucleates. In this regime spots travel downstream at approximately the mean velocity[7], U, and the survival rate of turbulence was determined a fixed distance downstream of the perturbation point. At each Reynolds number between 200 and 2000 events were investigated. In the case of duct flow the angle of the out flowing fluid at the duct exit was monitored to determine if the spot of turbulence has survived or decayed to laminar[10,11]. In the channel the pressure (with respect to ambient) was monitored 20 cm's from the channel exit using a commercial pressure sensor. The passage of a turbulent spot leads to a rapid fluctuation of the otherwise steady pressure signal. In all cases particular care was taken to ensure that the Reynolds number could be controlled to better than 0.2% for several hours. Such a good control is crucial to obtain reliable lifetimes. A detailed description of the temperature and flow rate control can be found in reference 11.

The expansion speed of spots was measured by monitoring the pressure drop over a length of ~250 L/D. Turbulence was first generated by a perturbation at a fixed location, 200 D upstream the measured section. This distance is long enough for turbulence to delete any memory from its creation so that the results are independent of the perturbation type[23]. When the turbulent spot arrives to the measured section the pressure drop increases due to the higher drag of the turbulent flow. Five stages can be identified during the passage of a turbulent spot: 1) laminar flow, 2) spot entrance, 3) spot inside, 4) spot exit and 5) laminar flow. Each of these stages is characterized by a different slope of the pressure trace. The front and rear velocities and expansion rates were determined by identifying the beginning of these stages in the pressure signal. For instance the front velocity is obtained by determining when the spot first enters ($t_{1\rightarrow2}$) and first exits ($t_{3\rightarrow4}$ ) the measured section and it is given by: $U_{FRONT} = L/(t_{3\rightarrow4} - t_{1\rightarrow2})$. At each Reynolds number expansion rates were averaged over 10-25 events.

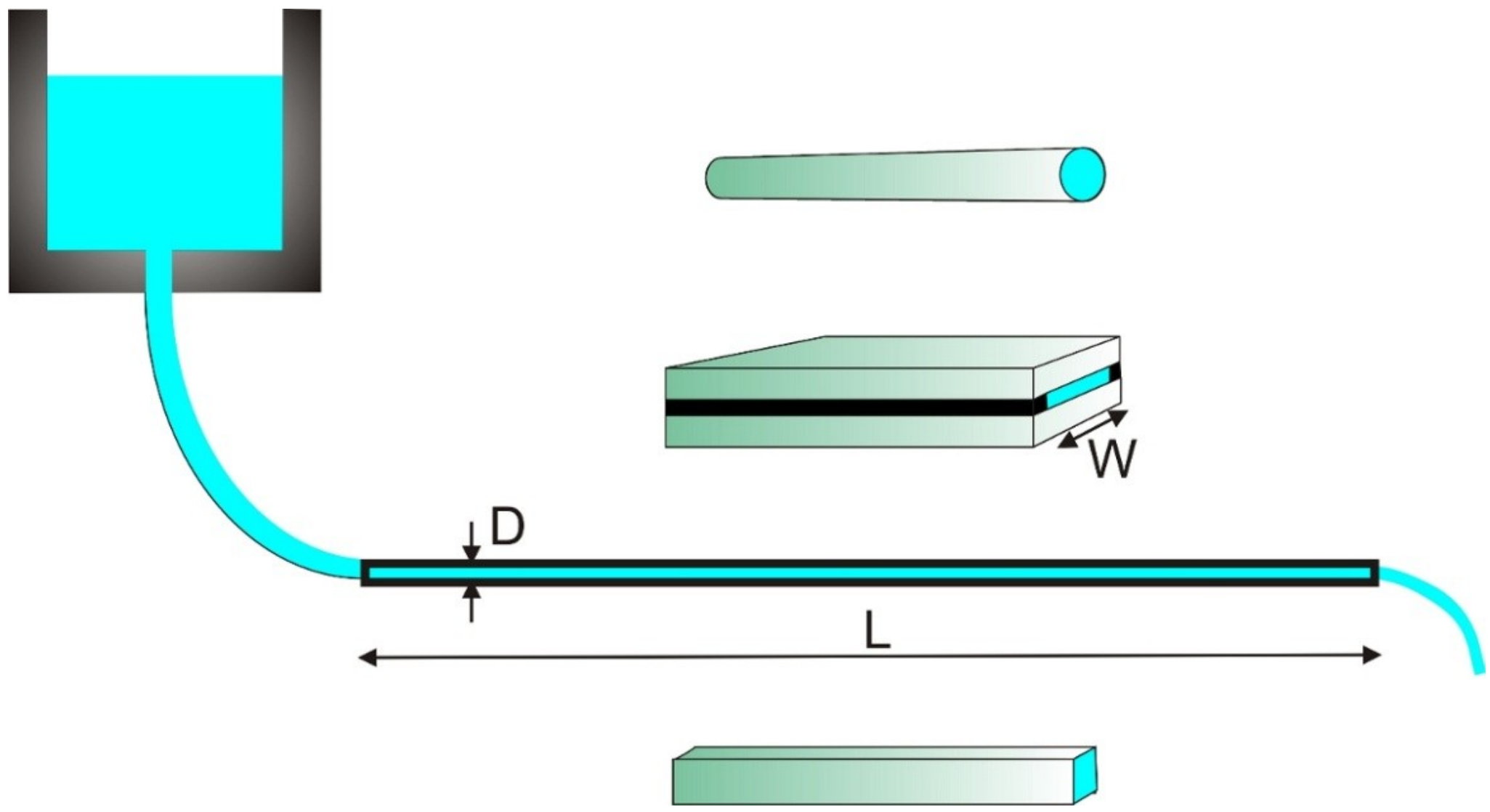

**Fig.1.** Schematic of the three experimental geometries. Experiments were carried out in a pipe of diameter D=30mm(±0.1) and a length of L=12m, a channel with a gap of D=4mm(±0.01), a width of W=120mm and length of L=3m and a 8m long square duct with a wall spacing of D=8mm(±0.1). The Reynolds number is defined using the duct/channel heights and the pipe diameter. Hence the dimensionless lengths (L/D) of the three facilities are 400, 750 and 1000 respectively while the channel aspect ratio (W/D) is 30.

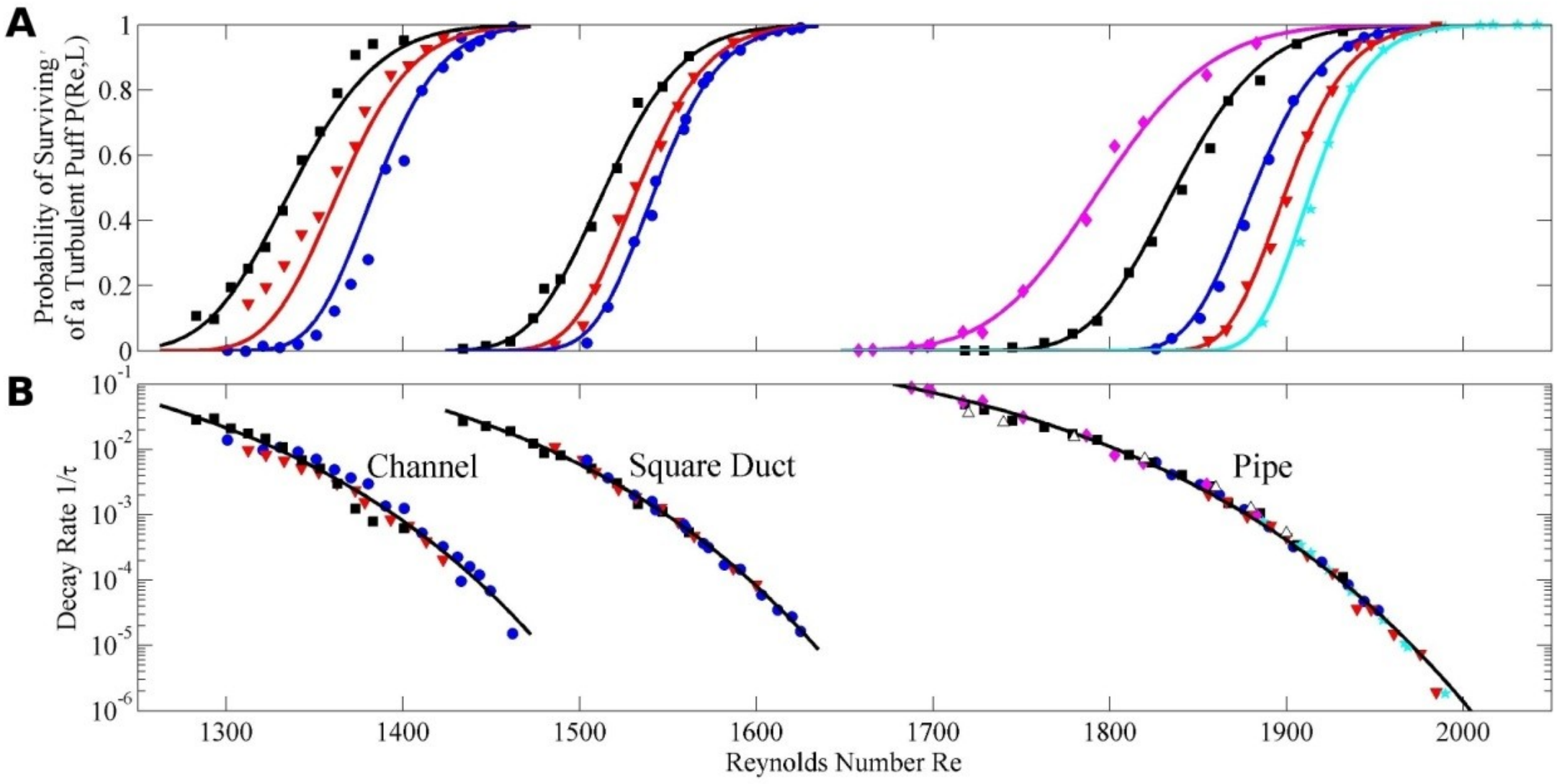

**Fig. 2.** Lifetimes of turbulent spots in channel, duct and pipe flow. **A** Probability of survival of a turbulent puff after travelling a distance L in three different flows. From left to right: channel, square duct and pipe. Different symbols correspond to different distances. In channel: squares (L=600D), triangles (L=362.5D) and circles (L=237.5D); in square duct: squares (L=875D), triangles (L=625D) and circles (L=372.5D); in pipe (same data as in reference 10): diamonds (L=143D), squares (L=269D), triangles (L=933D), circles (L=1900D) and stars (L=3450D). **B** Decay rate calculated as function of the Reynolds number from the data shown in A. The open triangles for the pipe correspond to the numerical simulations from reference 12. The lines (A and B) show the two parameter fit of the super exponential function $\tau^{-1} = \exp(-\exp(c[Re-Re_0]))$. Fitting parameters for the channel are c=0.0062, $Re_0$=1085, for the duct: c=0.0061, $Re_0$=1221 and for the pipe: c=0.0057, $Re_0$=1526.

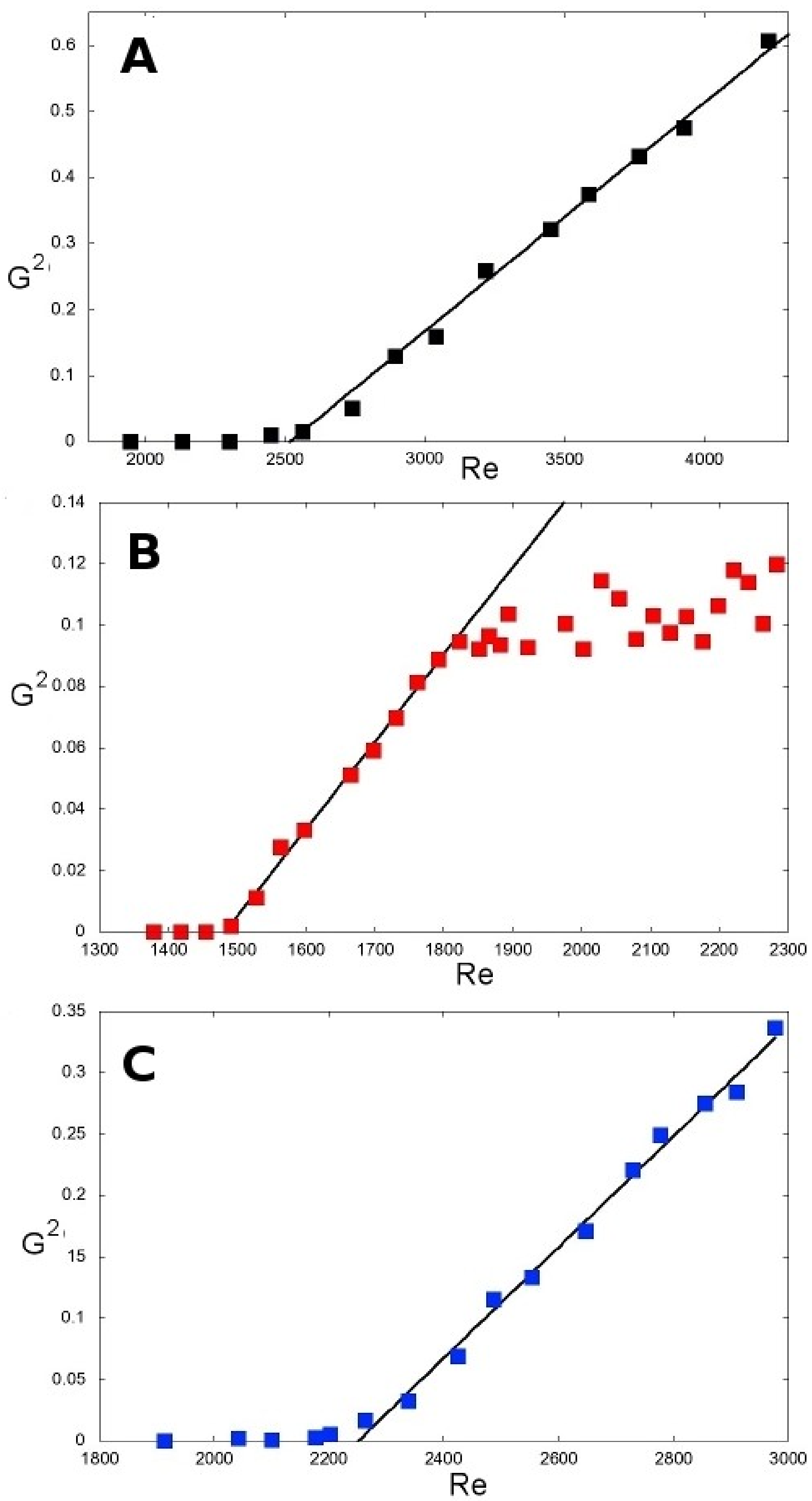

**Fig. 3.** Growth rate of a turbulent spot close to the phase transition. Square of the dimensionless growth rate of turbulent spots plotted as a function of Re close to the phase transition in pipe (A), channel (B) and square duct (C). The line follows $G \propto (Re - Re_C)^{1/2}$ which yields the following critical numbers: $Re_C^{pipe} = 2550$, $Re_C^{channel} = 1480$ and $Re_C^{duct} = 2250$. Note that for the channel non-linear saturation sets in much earlier than in the other two cases.